\documentclass[twocolumn,showpacs,pra,aps,floatfix,amssymb]{revtex4}
\usepackage{graphicx}
\usepackage{braket} 
\usepackage{float}  				    
\usepackage[utf8]{inputenc} 
\usepackage{natbib}
\usepackage{subfigure}
\usepackage{bm}
\usepackage{mathrsfs}

\usepackage{comment}
\usepackage{soul}
\usepackage{color}
 \RequirePackage{color}
 \definecolor{darkred}{rgb}{0.8,0.1,0.1}

\begin{document}

\newcommand{\beq}{\begin{equation}}
\newcommand{\eeq}{\end{equation}}
\newcommand{\beqa}{\begin{eqnarray}}
\newcommand{\eeqa}{\end{eqnarray}}

\title{Zitterbewegung and Klein-tunneling  phenomena for transient  quantum waves}
\author{Fernando Nieto-Guadarrama}
%
\author{Jorge Villavicencio}

\affiliation{Facultad de Ciencias, Universidad Aut\'onoma de Baja
California, 22800 Ensenada, B.C., M\'exico}
\email{villavics@uabc.edu.mx}
\date{\today}
%

\begin{abstract}

We explore the dynamics of relativistic quantum waves in a potential step by using an exact solution to the Klein-Gordon equation  with a point source initial condition. 
We show that in both the propagation, and Klein-tunneling regimes, the {\it Zitterbewegung} effect manifests itself as a series of {\it quantum beats} of the particle density in the long-time limit. 
We demonstrate that the beating phenomenon is characterized by the {\it Zitterbewegung} frequency,  and that the amplitude of these oscillations decays as $t^{-3/2}$.  
We show that beating effect also manifests itself in the free Klein-Gordon and Dirac equations within a quantum shutter setup, which involve the dynamics of cut-off quantum states.
We also find a time-domain where the particle density of the point source is governed by the propagation of a main wavefront, exhibiting an oscillating  pattern similar to the {\it diffraction in time} phenomenon observed in non-relativistic systems. 
The relative positions of these wavefronts are used to investigate the time-delay of quantum waves in the Klein-tunneling regime. 
We show that, depending on the energy difference, ${\cal E}$, between the source and the potential step, the time-delay can be  positive, negative or zero. 
The latter case corresponds to a super-Klein-tunneling configuration, where ${\cal E}$ equals to half the energy of the  potential step. 

\end{abstract}


\maketitle

\section{Introduction}

Transient phenomena are ubiquitous in the field of quantum dynamics where they have proven to be very useful tools to explore and manipulate the features of matter-waves\cite{k94, dcgcjm, mohsen2013quantum}.
Transients arise from the properties of quantum waves involving relativistic and non-relativistic equations subject to sudden changes in the initial condition. 
The most representative transient phenomenon is the {\it diffraction in time} of free matter-waves, predicted in 
Ref. \onlinecite{mm52} using a quantum-shutter setup that deals with cut-off initial states. 
The experimental observations of  diffraction in time were later realized in experiments with ultracold atoms \cite{dalibard}, cold-neutrons \cite{Hils98}, and atomic Bose-Einstein condensates \cite{colombe05}. 
The understanding and control of the features of diffraction in time are of relevance, for example, in the field of atom lasers \cite{Hagley99,tripp00,12290615}, which operate by extracting matter-waves from Bose-Einstein condensates.  
Some later theoretical works have also addressed the study of transients using various initial quantum states, as well as generalizations of the shutter model in order to explore different types of transients associated to the buildup process, resonance scattering, and tunneling dynamics in different potential profiles. For a review on the subject, see for example Ref. \cite{dcgcjm}. 
Moshinsky's pioneering  quantum shutter model deals also with the dynamics of free particles of spin 0 and 1/2, involving the Klein-Gordon \cite{mm52} and Dirac \cite{moshrmf52} equations, respectively. 
The dynamical features of relativistic quantum waves have also been explored 
with initial conditions other than the quantum shutter setup.
This is the case of the \textit{point source} initial condition problem that has been of great interest since the original  proposal  by  Stevens \cite{Stevens_1983} of  tunneling monochromatic fronts. Aside from the controversy regarding the propagation of such tunneling wavefronts \cite{Moretti_1992,teranishi,JAUHO1989303}, the issue was later investigated in Refs. \onlinecite{mbmuga00} and \onlinecite{jvrrsss02},
that showed the existence of non-tunneling forerunners in evanescent media, with the interesting result that the time of arrival of these transient structures is governed by the traversal or  B\"{u}ttiker-Landauer time \cite{PhysRevLett.49.1739}.
These works were developed in the context of Schr\"{o}dinger's equation, and later inspired the study of relativistic transient forerunners with evanescent conditions using a Klein-Gordon type equation \cite{dmrgv03}.
Although in these works the study of precursors and their relevant time scales have been of great interest in the problem of time-dependent features of wave propagation in evanescent media, there are other aspects predicted by quantum electrodynamics that have been not been fully explored (or overlooked) using a relativistic point source model. This is the case, for example, of the  transient features of quantum waves in the Klein-tunneling regime, and the {\it Zitterbewegung} (ZBW) effect.
Interestingly, these types of phenomena originally predicted in the context of quantum electrodynamics, have 
received a renewed interest due to the prediction of such relativistic effects in 2D material such as graphene \cite{Novoselov666}.
In fact, the ZBW effect \cite{Katsnelson2006, Krueckl_2009}, and Klein-tunneling have now become accessible to experiments \cite{Klein3,Gerri}, 
and also the importance of transient nature of the ZBW has been stressed out \cite{PhysRevB.76.195439}. 
Theoretical models dealing with Gaussian wavepacket dynamics have proven to be  powerful methods to study transient quantum wave dynamics and ZBW phenomena in physical systems such as semiconductors \cite{Zawadzki_2011,Schli_05,PhysRevB.78.115401,PhysRevB.72.085217}, monolayer and bilayer graphene \cite{PhysRevB.78.235321,martinez10,PhysRevB.76.195439,PhysRevB.80.045416,Krueckl_2009,PhysRevB.80.165416}, silicene \cite{ROMERA20142582,siliceno2}, and phosphorene \cite{PhysRevB.99.235424}.
Recently, a signature of the ZBW effect in the particle density has been reported in a model involving the dynamics of a massive Dirac particle in the vicinity of a black hole \cite{Dirac-Curved}, using an initial Gaussian wave. 

In this work we propose an alternative approach  of addressing  ZBW  phenomena by exploring the relativistic dynamics using a point source model. We
study the transient features of relativistic quantum waves by using a time-dependent solution to the Klein-Gordon equation for a potential step with a point source initial condition.
Our aim is to show that the ZBW effect reveals itself in the long-time behavior of the particle density as a series of {\it quantum beats}, and that this feature provides us with a useful alternative tool to explore the ZBW.
We also characterize the dynamical features of the propagation and Klein-tunneling regimes for different source energies. 
Our work is organized as follows: in Sec.~ \ref{model} we present the main features of the relativistic point source model for a potential step, and set the main equations. In Sec.~\ref{QWE} we explore the  dynamics of quantum waves, and finally in Sec.~\ref{CONCLU} we present the conclusions. 

\section{Point source model for a potential step} \label{model}

Let us consider the one-dimensional Klein-Gordon equation for a scalar
potential $U(x)$ in the Coulomb-Gauge,
\beq
\partial^2_x \psi(x,t)=\mu^2 \,\psi(x,t)-\left[\frac{i}{c} \partial_t-\frac{U(x)}{\hbar c}\right]^2\psi(x,t),
\label{kg_equation}
\eeq
where $\mu=mc/\hbar$, for a  step potential $U(x)$ defined as
\beq
U(x)=\cases{U_r, &  $x\ge 0;$\cr 0, &  $x < 0$,}
\label{eq1}
\eeq
with constant $U_r>0$.
Our choice of  Klein-Gordon equation to study transient relativistic phenomena of spin $0$ particles, is for simplicity. 
This is because in the case of Dirac particles of spin $1/2$, 
the spin is irrelevant \cite{mm52} in the dynamics when choosing a fixed initial spin direction. Thus, it is expected that the main features observed in the time-evolution of relativistic spinless systems, also manifest themselves in systems with spin $1/2$.
The point source problem for a step potential barrier involves the solution of Eq.~(\ref{kg_equation}) with the initial condition of a source with sharp onset, namely,
\beq
\psi(0,t)=\cases{e^{-i E c t}, &  $t>0;$\cr 0, &  $t < 0$,}
\label{CI}
\eeq
that follows from the general stationary solution to Eq.~(\ref{kg_equation}) given by $\psi(x,t)=e^{i(k x- E c t)}$, with a dispersion relation 
\beq
 (E-V)^2= k^2+ \mu^2,
\label{dispersion_kg_norm} 
\eeq
where $E=(E_r/\hbar c)$, and $V=(U_r/\hbar c)$, are the energies in units of the reciprocal length, with  
$k$ the corresponding momentum.
To obtain an exact time-dependent solution for $x>0$ and $t>0$, we propose a solution of the form
\beq
\psi(x,t)=\psi_0(x,t)\,e^{-i V c t},
\label{psit_traslada}
\eeq
that allows to rewrite Eq.~(\ref{kg_equation}) as a free-type equation,
\beq
\partial^2_x \psi_0(x,t)=\mu^2 \,\psi_0(x,t)+\frac{1}{c^2}\,\partial^2_t \psi_0(x,t),
\label{kg_equation_2}
\eeq
with the initial condition,
\beq
\psi_0(0,t)=\cases{e^{-i {\cal E}c t}, &  $t>0;$\cr 0, &  $t < 0$,}
\label{CI_new}
\eeq
with ${\cal E}=(E-V)$ defined as the energy of an equivalent free-type source.
%
The  time-dependent solution $\psi_0(x,t)$  can be obtained from Eq.~(\ref{kg_equation_2}) with the initial condition given by Eq.~(\ref{CI_new}), by following the procedure discussed in Refs.  \onlinecite{dmrgv03} and \onlinecite{jpajv00}. For completeness, in Appendix \ref{PSB} we discuss the solution method for the free-type case. 
Thus, the relativistic solution for the point source problem for a potential step reads,
\beq
\psi(x,t)=\left\{ 
\begin{array}{ll}
\psi_+(x,t)+\psi_-(x,t), & t>x/c; \\ 
0, & t<x/c,
\end{array}
\right.
\label{stepfinal}
\eeq
where we have defined  $\psi_{\pm}(x,t)$ as
\beqa
\psi_{\pm}(x,t)&=& \left[e^{i[\pm kx-{\cal E}ct]}+\frac{1}{2}%
J_0(\eta ) \right. \nonumber \\
&&\left. -\sum\limits_{n=0}^\infty (\xi /iz_{\pm })^nJ_n(\eta ) \right]\,e^{-i c Vt}.
\label{simplifbis2}
\eeqa
An alternative representation of 
Eq. (\ref{simplifbis2}) can be obtained with the help of the Bessel generating function  $e^{\frac{\nu}{2}(z-z^{-1})}=\sum_{m=-\infty}^{\infty}z^m J_m(\nu)$,
that leads us to an expansion 
\beqa
\sum\limits_{n=0}^\infty \left(\frac{\xi}{iz_{\pm }}\right)^n J_n(\eta)&&= e^{i[\pm k x-{\cal E}c t]}
 \\
&&-\sum\limits_{n=1}^\infty (-1)^n \left(\frac{iz_\pm}{\xi}\right)^n J_n(\eta), \nonumber
\eeqa
so, Eq. (\ref{simplifbis2}) may also be written in the form 
\beq
\psi_{\pm}(x,t)= \left[\frac{1}{2}J_0(\eta )
+\sum\limits_{n=1}^\infty (-1)^n (iz_\pm/\xi)^n J_n(\eta )\right]e^{-i c Vt}, 
\label{pre}
\end{equation}
which will be useful for the analysis of the
region close to the relativistic cutoff, and also for the long-time behavior of the relativistic quantum wave.
We shall use solution (\ref{stepfinal}) to calculate the particle density $\rho(x,t)$ defined as
\beq
\rho(x,t)=-\frac{\hbar}{mc^2}{\rm Im}\left[ \psi^*\partial_t\psi \right] -\frac{ U_r }{m c^2}\,\psi^*\psi, \nonumber\\
\label{currdens_natural_2} 
\eeq
in order to explore its main transient features  as  function of time and position in the  different energy regimes.

\section{Quantum wave evolution}\label{QWE}
\begin{figure}[H]
{\includegraphics[angle=0,width=3.2in]{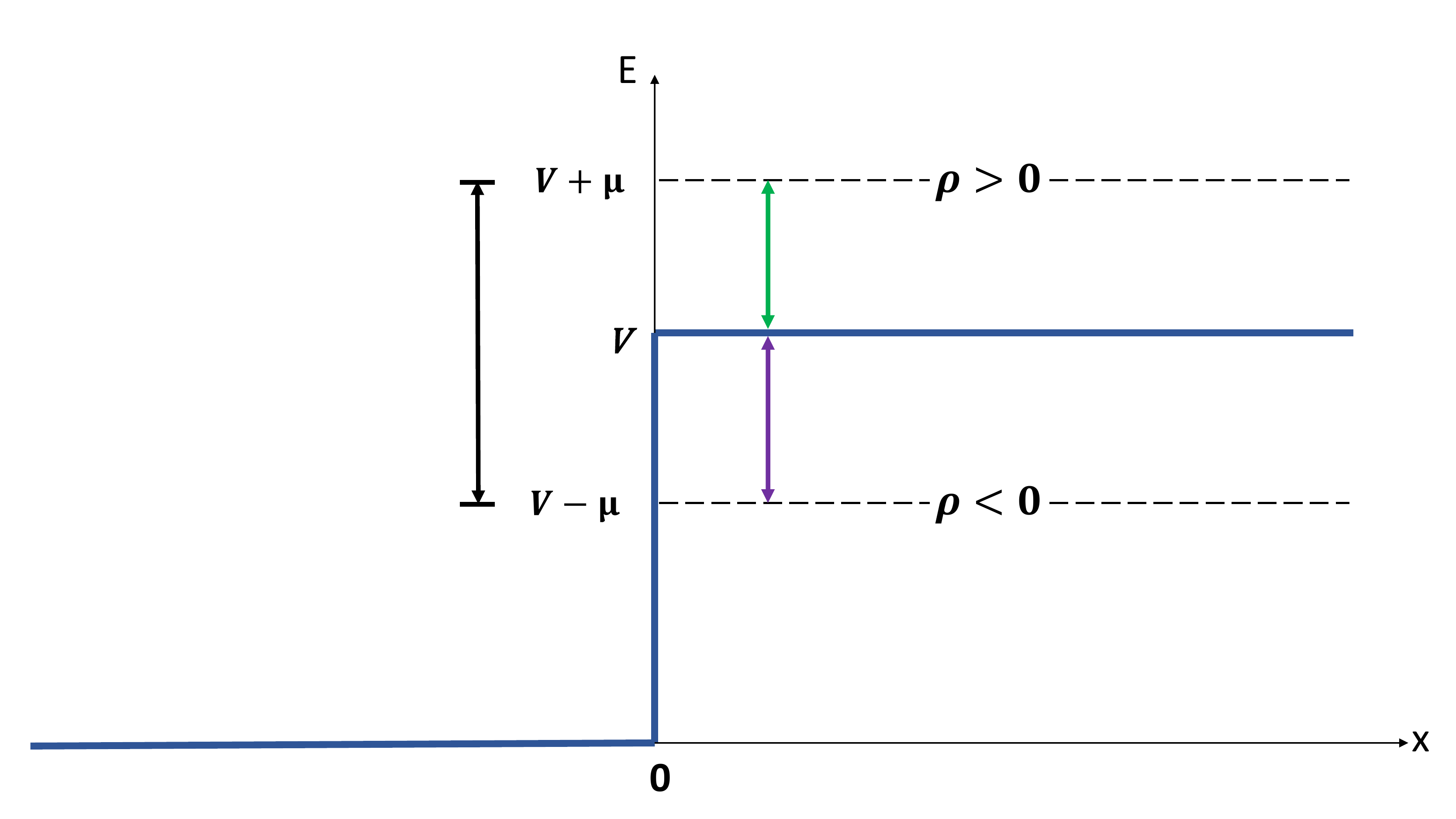}}
\caption{Step potential of intensity $V$, and the energy range of the different regimes corresponding to: (i) propagation regime ($E>V+\mu$),  (ii) evanescent case ($V-\mu<E<V+\mu$),  and (iii) Klein-tunneling regime ($E<V-\mu$). Here and in all of our numerical calculations we use $\hbar=m=c=1$. }
\label{wavefr}
\end{figure}

We use the solution $\psi(x,t)$ given by Eq.~(\ref{stepfinal}) to explore the particle density $\rho(x,t)$ [Eq.~(\ref{currdens_natural_2})], as  function of position $x$ and time $t$,  
for a potential step of intensity $V$, and a source energy, $E$. 
From the dispersion relation given by Eq.~(\ref{dispersion_kg_norm}), the momentum $k$ deals with real or imaginary values, which defines the different regimes of interest.
These regimes are governed by the dispersion relation  Eq.~(\ref{dispersion_kg_norm}), corresponding to \textit{propagation} ($E>V+\mu$) with real positive values of momentum $k=+\sqrt{{\cal E}^2-\mu^2}$,  the \textit{evanescent} case ($V-\mu<E<V+\mu$) involving imaginary values of momentum $k=i\,\sqrt{\mu^2-{\cal E}^2}$,  and the \textit{Klein-tunneling regime} ($E<V-\mu$) with negative values of momentum  $k=-\sqrt{{\cal E}^2-\mu^2}$. See Fig.~\ref{wavefr}.
In sections \ref{Spropaga}, and \ref{SKlein}, we will first focus on characterizing the time-dependent features of quantum waves with real values of $k$, namely, the  propagation and Klein-tunneling regimes, respectively. In section \ref{SZitter} we study and characterize the ZBW phenomenon in these regimes of interest, by exploring the long-time behavior of the particle density. 
In Sec.~\ref{relativistic_shutter_models} we show that the \textit{quantum beat} phenomenon of the ZBW effect also manifest itself for relativistic particles of spin $0$ and $1/$2, within a \textit{quantum shutter model} involving the free Klein-Gordon \cite{mm52} and Dirac equations \cite{moshrmf52}.   
\begin{figure}[!tbp]
{\includegraphics[angle=0,width=3.4in]{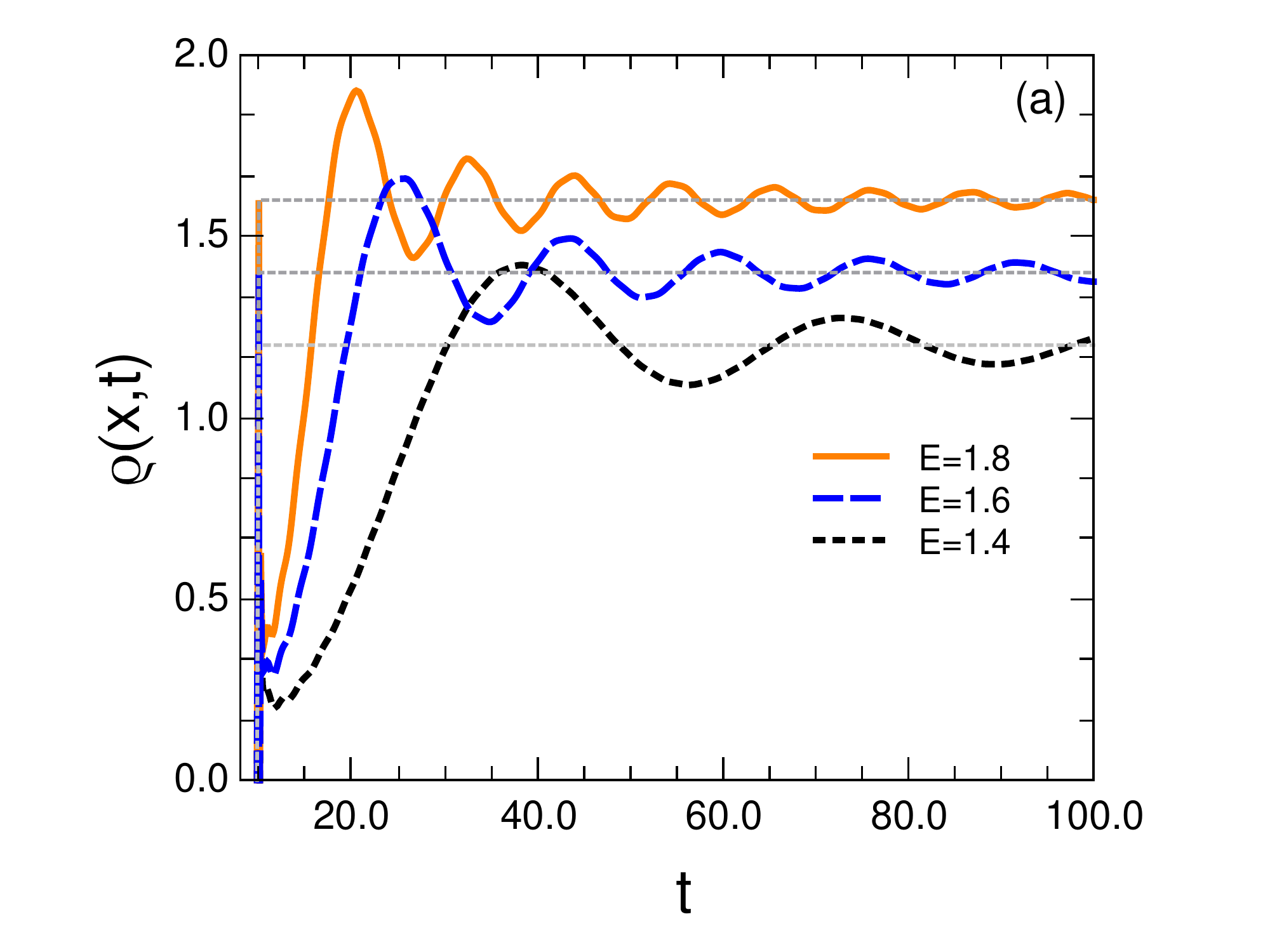}
\includegraphics[angle=0,width=3.4in]{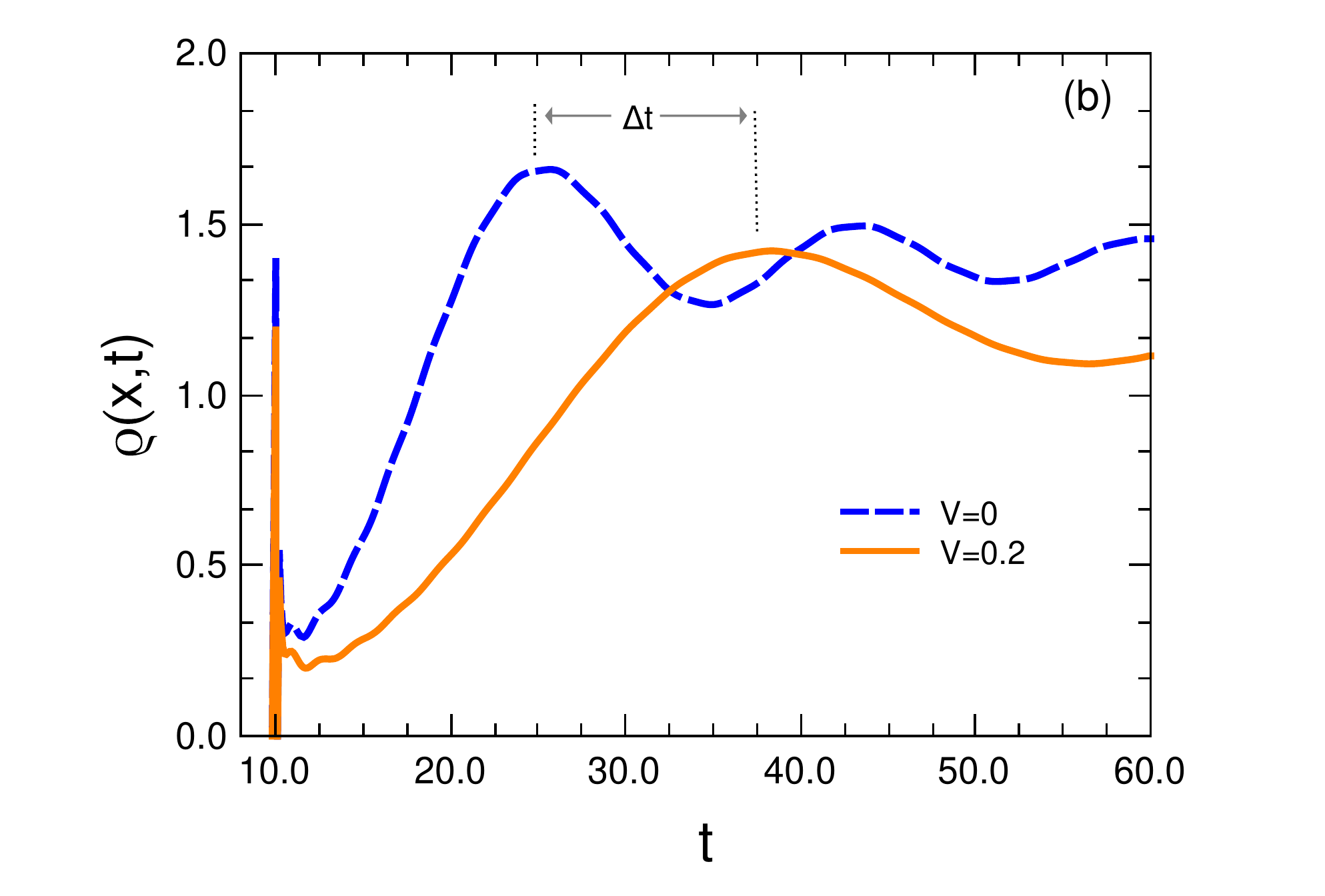}}
\includegraphics[angle=0,width=3.35in]{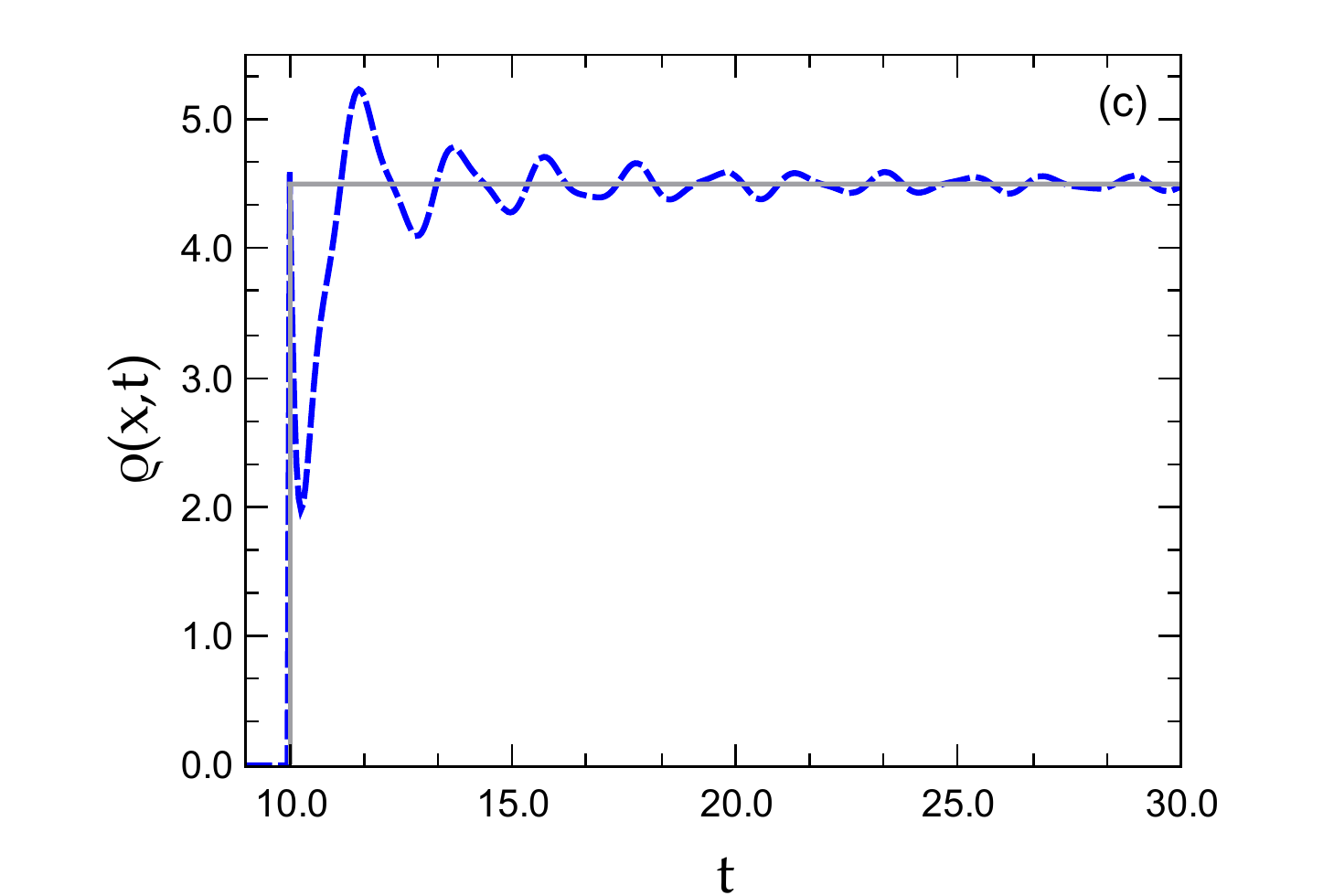}
\caption{(a) Time-evolution of the particle density $\rho$ [Eq.~(\ref{currdens_natural_2})] at the position $x=10.0$ for  $V=0.2$, and different values of the source energy, $E$. The stationary density $\rho={\cal E}$ (grey dotted line) is included for comparison in all cases. The density exhibits an oscillating pattern similar to {\it diffraction in time} phenomenon.  The velocity of the main wavefronts increases as the energy of the source $E$ (or the free-type energy ${\cal E}$) increases. (b) Dynamical delay $\Delta t$ of the free-type density (orange solid line) with respect to the free point source case ($V=0$) (blue dashed line) for the case $E=1.6$. (c) Time-evolution of $\rho$ at the position $x=10.0$ for high energy, $E=5$ and $V=0.5$, where we can see a distortion respect with the time-evolution shown in case (a). The stationary solution (gray solid line) is also included.}
\label{kg_propaga_a_new}
\end{figure}

\subsection{Propagation regime}\label{Spropaga}
The propagation regime corresponds quantum waves with  real positive values of momentum (see Fig.~\ref{wavefr}),
and is similar to Schr\"odinger's step potential case \cite{jvrrsss02}, where the energy is above the step. We consider in Fig. \ref{kg_propaga_a_new}(a)  the time-evolution of $\rho(x,t)$, where the density exhibits a sharp relativistic wavefront at a time $t_F=(x/c)$. The dynamics of this relativistic front  
is governed by the solution (\ref{stepfinal}) using Eq.~(\ref{pre}), wherein the limit $x \to ct$ the arguments of the Bessel functions $J_{\nu}(\eta)$ are very small ({\it i.e.} $\eta\simeq0$). Therefore,  by using the asymptotic form of the Bessel function for small arguments \cite{abramowitz1964handbook}, $J_{\nu}(\eta)\simeq(\eta/2)^{\nu}\,(\nu!)^{-1}$, and identifying the resulting exponential series expansions, we obtain the solution,
\beq
\psi(x,t)\simeq \left[-1+e^{ -i\mu z_+\,(ct-x)/2 }+e^{-i \mu z_-\,(ct-x)/2} \right ] e^{-i c Vt}, \label{asymsol}
\eeq
used in our calculation of the density in the vicinity of the relativistic wavefront. 
In all cases discussed in Fig. \ref{kg_propaga_a_new}(a), for values of $t_F$ onward, $\rho(x,t)$ grows towards a maximum value from which it oscillates until it reaches the stationary value, governed by the plane-wave  $\psi_s\rightarrow e^{i(kx-Ect)}$. 
The corresponding stationary density $\rho={\cal E}$ is included for comparison in Fig.~\ref{kg_propaga_a_new}(a). 
The oscillations of $\rho(x,t)$ in Fig. \ref{kg_propaga_a_new}(a) resemble the {\it diffraction in time} effect predicted in Ref. \onlinecite{mm52} for the case of non-relativistic free matter-waves.
The time-diffraction effect is an oscillatory pattern exhibited by the probability density within a quantum shutter setup involving cut-off plane waves  \cite{mm52,mm76}, resembling the optical diffraction of light by a semi-infinite plane.  
Interestingly, in our relativistic case, our  solutions $\psi_{\pm}(x,t)$ [Eq.~(\ref{simplifbis2})]  can be expressed in terms of Lommel functions of two variables, originally introduced in the context of optical diffraction problems \cite{Watson96}. 
Moreover, we can appreciate in Fig. \ref{kg_propaga_a_new}(a) that the main wavefront propagates along the structure with a velocity that is proportional to $E$ (or ${\cal E}$). Also, the main wavefronts associated with higher energies are faster than the wavefronts with lower energies.
In Fig.~\ref{kg_propaga_a_new}(b) we compare the densities for the free-type case and the free point source ($V=0$). 
We show that the density associated to the free-type case, exhibits a dynamical time-delay $\Delta t=(t_0-t_V)$, obtained by measuring the difference of the position of the maximum values of $\rho(x,t)$ 
at $t_0$ and $t_V$, corresponding  to the free case and the free-type source, respectively. 
For the propagation regime the energy of the free-type source,  ${\cal E}$, is always smaller than the energy  of the free source, $E$, that is, ${\cal E}<E$, and as a consequence, a time-delay (positive delay) is always observed. 

The general features of relativistic quantum waves in the propagation regime Fig.~\ref{kg_propaga_a_new} can be summarized as follows.
A precursor associated to the quantum wave propagates at a velocity $c$, which characterizes the arrival of an early signal. At later times, this precursor is then followed by a main wavefront, exhibiting an effect similar to {\it diffraction in time}. 
The main wavefront of the particle density always exhibits a positive time-delay ($\Delta t>0$) in the propagation regime. 
There is another regime where the momentum $k$ is also real. This is the so-called Klein-tunneling regime that we shall study in the next section with the help of the results obtained in the present section.

\subsection{Klein-tunneling regime \label{SKlein}}

\begin{figure}[H]
{\includegraphics[angle=0,width=3.3in]{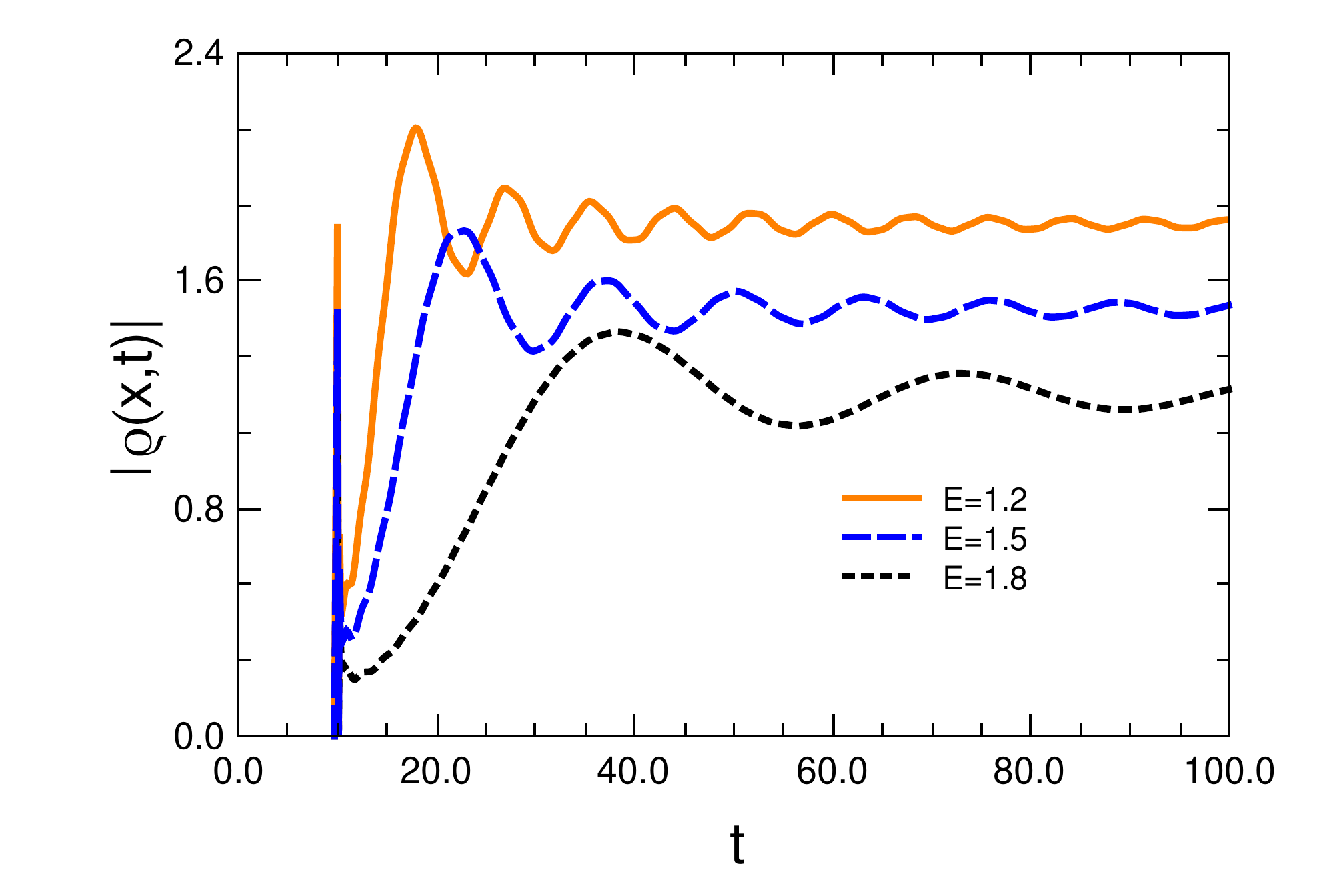}}
\caption{Time-evolution of the particle density $|\rho(x,t)|$ [Eq.~(\ref{currdens_natural_2})] in the Klein-tunneling regime for $V=3.0$ at a fixed value of position $x=10.0$, for different values of the source energy, $E$. }
\label{kg_klein_ALLE}
\end{figure}

The Klein regime is characterized by a source energy below the potential step, with a real negative value of momentum (see Fig.~\ref{wavefr}), that has no non-relativistic \cite{jvrrsss02} counterpart. 
In Fig.~\ref{kg_klein_ALLE} we analyze the time-dependent features of the density in the Klein regime. Since $\rho<0$, and also to simplify the comparison with the free source case, we plot $|\rho(x,t)|$ to help the eye.
We observe that, although  the energy of the source is below the step, the density exhibits features typical of quantum wave propagation, as well as the {\it diffraction in time} transient, as discussed in Sec.~\ref{Spropaga}. 
However, the observed result in the Klein regime that quantum waves with a lower source energy $E$ are faster than those with lower energies, apparently contradicts the results of Sec.~\ref{Spropaga}.
\begin{figure}[H]
{\includegraphics[angle=0,width=3.4in]{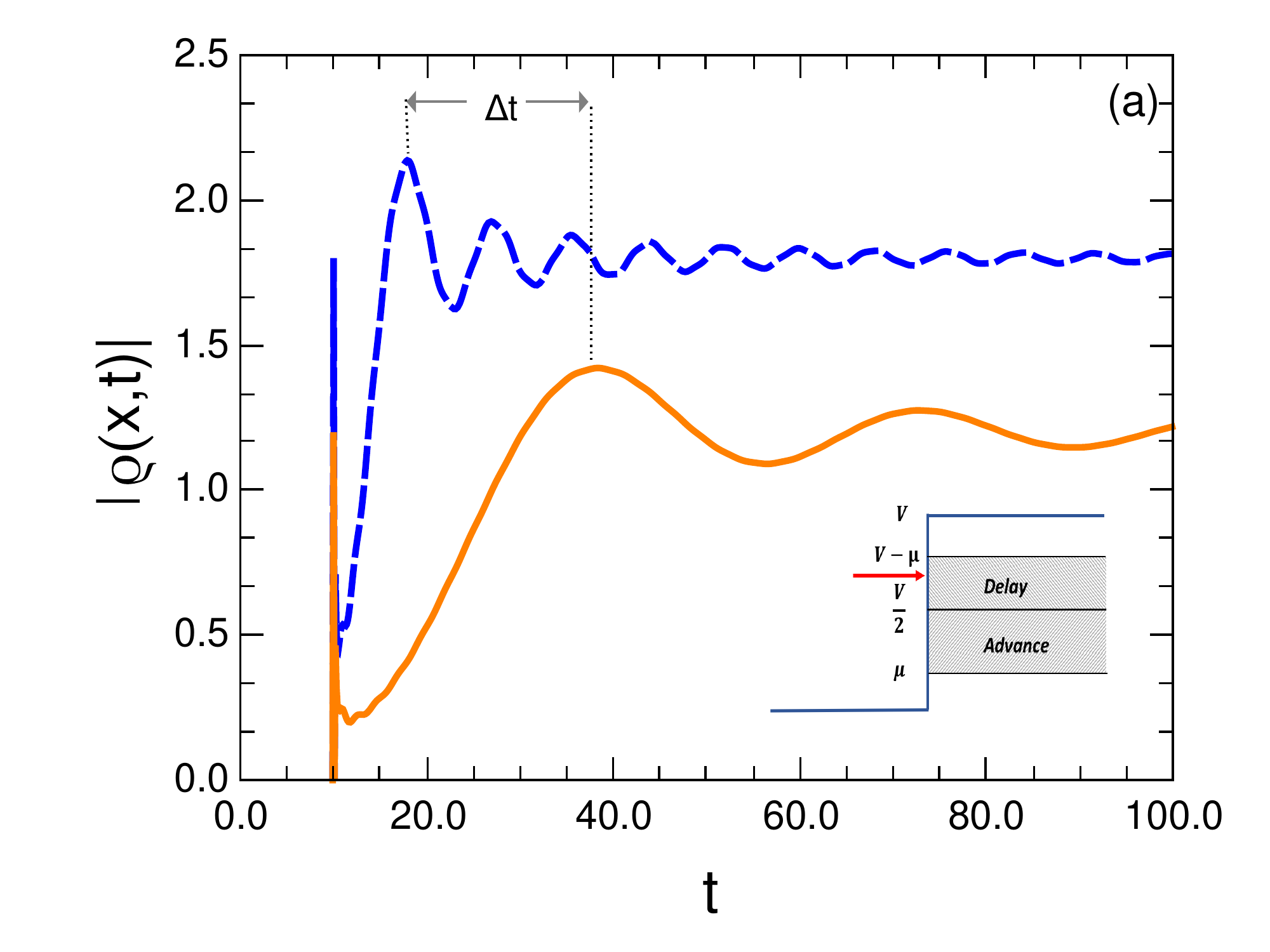}
\includegraphics[angle=0,width=3.4in]{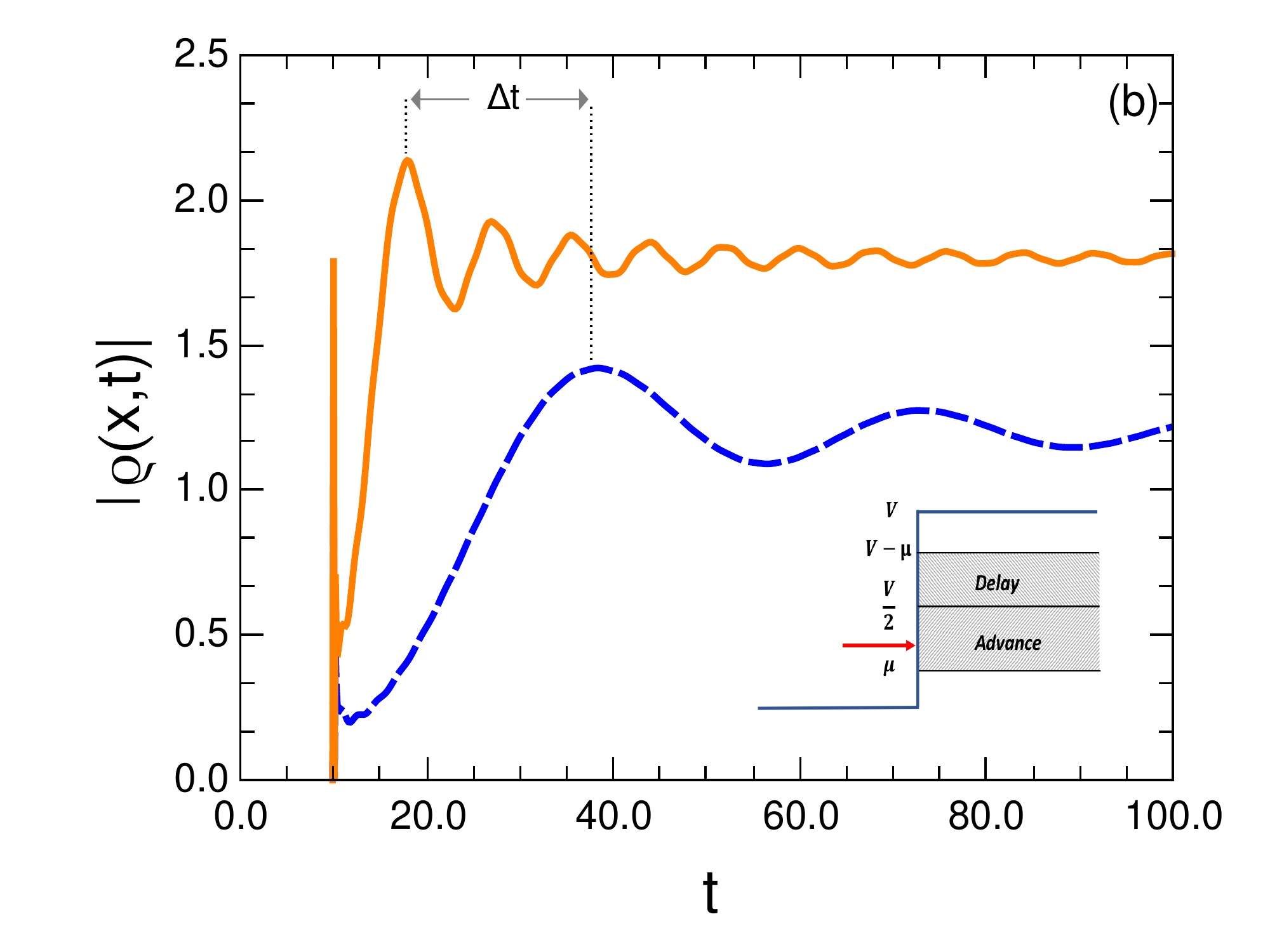}
\includegraphics[angle=0,width=3.4in]{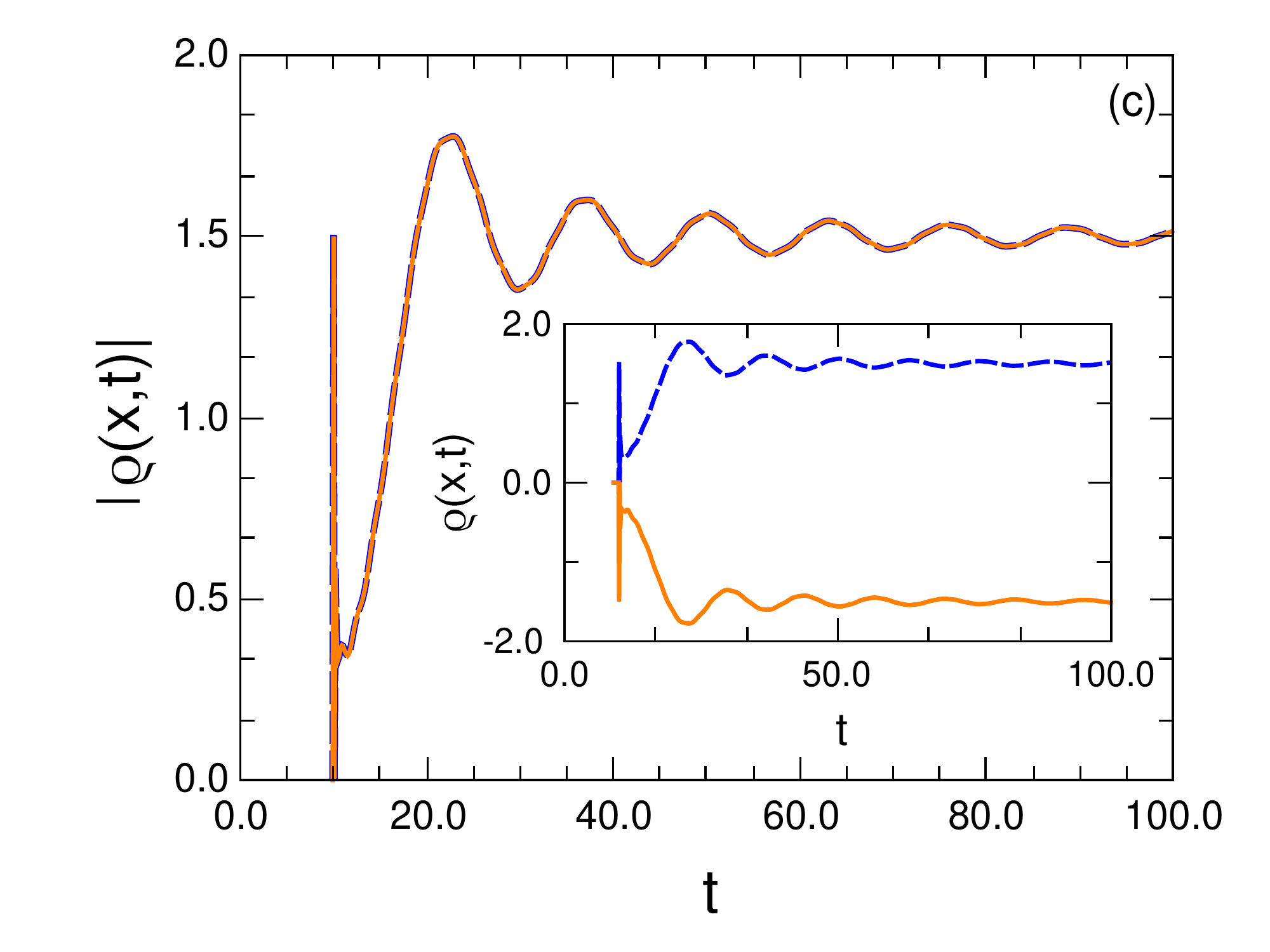}}
\caption{Time evolution in the Klein-tunneling regime of the particle density $|\rho(x,t)|$  [Eq.~(\ref{currdens_natural_2})] at a fixed position $x=10.0$ for a potential step with $V=3.0$ (orange solid line) for different values of the source energy, $E$, and compared with the free case  ($V=0$) density (blue dashed line). (a) For an  energy $E=1.8$ we observe a positive delay (time-delay), while in case (b) with $E=1.2$ a negative delay (time-advance) is observed. In case (c) a source energy $E=1.5$  yields $\Delta t=0$. See the inset for the comparison of the positive and negative densities, $\rho(x,t)$. }
\label{Klein_allgoes}
\end{figure} 
Nevertheless, we shall see that there is no contradiction at all if we analyze the behavior of the wavefronts in terms of the free-type energy of the source, ${\cal E}$ instead of $E$, as we discuss below in the context of the dynamical time-delay.
In this respect, we observe a peculiar effect of the dynamical delay $\Delta t$ associated to the particle density. %
In Fig.~\ref{Klein_allgoes},  we compare the time of arrival the main wavefront of the density associated to the potential with respect to that related to the free case.
In Fig.~\ref{Klein_allgoes}(a) and Fig.~\ref{Klein_allgoes}(b) we observe that the dynamical delay can be positive (time-delay) or negative (time-advance), respectively.
For example, in the case of  Fig.~\ref{Klein_allgoes}(a) (Fig.~\ref{Klein_allgoes}(b)) we can appreciate that
$|{\cal E}|<E$ ($|{\cal E}|>E$)
so the propagation of a faster (slower) main wavefront gives rise to a time-delay (time-advance).
Interestingly, in the case of  $|{\cal E}|=E$ shown in Fig.~\ref{Klein_allgoes}(c), 
no time delay or advance is observed due to the same velocity of both wavefronts.  
We have found that the dynamical time-delay $\Delta t$ for the particle density in the Klein regime exhibits a time-delay for $V/2<|{\cal E}|<V-\mu$ or a time-advance for $<\mu<|{\cal E}|<V/2$, as long as the condition for $V>2\mu$ is fulfilled.  
For the case $|{\cal E}|=V/2$, $\Delta t=0$ {\it i.e.} there is no time-delay or time-advance observed for the particle density. Interestingly, in the stationary dispersion problems, the incidence condition $|{\cal E}|=V/2$ (or incidence at $E=V/2$) gives rise to the so-called {\it super-Klein tunneling} \cite{KIM20191391}.

\subsection{Zitterbewegung and the particle density}\label{SZitter}
\begin{figure}[H]
{\includegraphics[angle=0,width=3.3in]{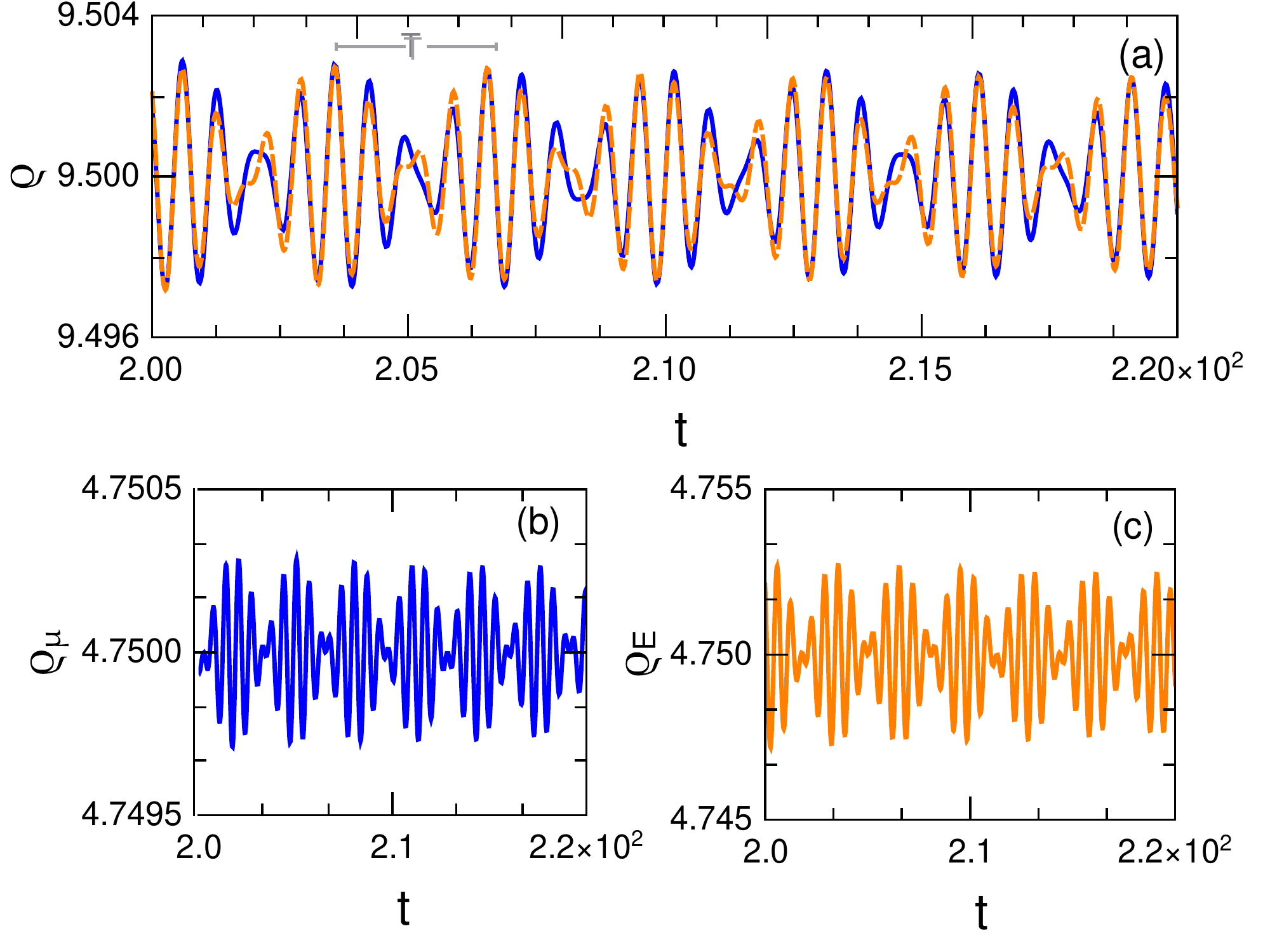}}
\caption{ (a) The ZBW effect manifests in the solution of the point source model a series of \textit{quantum beats} in the long-time behavior of particle density calculated from Eq.~(\ref{stepfinal}) (blue solid line) with an energy $E=10.0$, and $V=0.5$ for a fixed value of position $x=10.0$. The asymptotic density $\rho_a$ (orange dashed line) reproduces the transient behavior. The exact period of the beatings, $T = \pi$, is indicated in the density plot. The contributions to the asymptotic  $\rho_a$ given by (b) $\rho_{\mu}$  (blue solid line), and (c) $\rho_{\cal E}$ (orange solid line) exhibit a beating effect governed by the ZBW frequency, $\Omega$.}
\label{kg_zb_1}
\end{figure}

The ZBW effect \cite{PhysRevD.23.2454} (trembling motion) is due to the interference between the positive and negative-energy solutions of relativistic equations. The high-frequency oscillations of  free-Dirac particles are governed by a frequency $\Omega=(2 m c^2/\hbar)$, of the order of $10^{21}$ Hz, not accessible by present experimental techniques. 
An alternative approach to experimentally explore the ZBW is by implementing simulations of relativistic quantum effects within different physical setups. 
The pioneering work in solids by Schliemann \cite{Schli_05} and Zawadzki \cite{PhysRevB.72.085217} demonstrated that semiconductor electrons experience a ZBW effect, by establishing an analogy between the band structure of narrow-gap semiconductors and the Dirac equation for electrons in vacuum.   
Since then, experimental observations of ZBW phenomenon have been performed in simulations involving two-dimensional sonic crystal slabs \cite{PhysRevLett.101.264303}, 
optical superlattices \cite{PhysRevLett.105.143902}, trapped ions \cite{Gerri}, and Bose-Einstein condensates \cite{LeBlanc_2013}.
Other interesting  proposals on measuring ZBW deal with new materials like graphene \cite{Novoselof05},  where circularly polarized light \cite{PhysRevB.91.075419} is used to create a semiconductor-like spectrum in monolayer graphene, creating a system similar to the narrow-gap semiconductors of Ref.~\onlinecite{PhysRevB.72.085217}, which are known to exhibit the ZBW.   
On the theoretical side, the ZBW is explored by means of the time-evolution in Heisenberg's picture of the expectation value of the electrons position for wave packets \cite{Schli_05,PhysRevB.78.235321,PhysRevB.78.115401,Krueckl_2009,martinez10,Zawadzki_2011,PhysRevB.80.045416,PhysRevB.80.165416,ROMERA20142582,siliceno2,PhysRevB.99.235424}.  
We propose an alternative time-dependent approach to address the issue of ZBW, by exploring the features of the probability density at very long-times ($t\gg t_F$), which allows us to describe the dynamics using simple asymptotic formulas. 
In Fig.~\ref{kg_zb_1}(a) we consider the time-evolution of $\rho(x,t)$,  which exhibits high-frequency oscillations around the stationary value, 
similar to a superposition of quantum waves with different frequencies.
To identify the underlying  quantum superposition, we derive an asymptotic formula for the particle density $\rho_a(x,t)$ at very long times for the propagation regime.
We have found that the solution for $t\gg t_F$ is $\psi(x,t)\equiv\psi^{P}(x,t)=\psi_{+}^{P}(x,t)+\psi_{-}^{P}(x,t)$, with
\beqa
\psi_{+}^{P}(x,t)&&\simeq\left\{e^{i[kx-(E-V)ct]}-\frac{1}{2}J_0(\mu ct)
 \right. \nonumber \\ 
 && +i \left. \left(\frac{\xi}{z_+} \right) J_1(\mu ct)\right\}e^{-icVt},
\label{Psiasmas}
\eeqa
and
\beq
\psi_{-}^{P}(x,t)\simeq\left[\frac{1}{2}J_0(\mu ct)-i \left(\frac{z_-}{\xi} \right) J_1(\mu ct)\right]e^{-icVt},
\label{Psiasmen}
\eeq
thus, the solution for large values of  $t$ is
\beq
\psi^{P}(x,t)\simeq\psi_s^+(x,t)+i \left(\frac{\xi}{z_+}-\frac{z_-}{\xi} \right) J_1(\mu ct)e^{-icVt},
\label{psilt}
\eeq
where $\psi_s^+(x,t)=e^{i(k x- E c t)}$ stands for the stationary solution.
To obtain the long-time leading contributions of Eq.~(\ref{psilt}), we perform some algebraic manipulations in the prefactor of the Bessel function in Eq.~(\ref{psilt}), and  use the principal asymptotic form of the Bessel function \cite{abramowitz1964handbook}  $J_{\nu}(y)\simeq (2/\pi y)^{1/2}[\cos(y-\pi\nu/2-\pi/4)]$,
to obtain,
\beqa
\psi^{P}(x,t)\simeq\psi_s^+(x,t)  
+\frac{i\alpha}{z_+}\,\frac{e^{-icVt}}{t^{3/2}}\, \cos\left[\mu ct-\frac{3\pi}{4}\right],
\label{psiasym}
\eeqa
with $\alpha= (2/\pi \mu c)^{1/2}(2x/c)$. 
Note that in the limit $t \rightarrow \infty$, that the time-dependent solution tends to the stationary case {\it i.e.}  $\psi^{P}(x,t)\rightarrow \psi_s^+(x,t)$, as expected in the propagation regime.
Thus, the particle density $\rho_a(x,t)$ for the propagation regime is given by $\rho_a\simeq\rho_{{\cal E}}+ \rho_{\mu}$ with
\beqa
\rho_{{\cal E}}&\simeq&\frac{\hbar}{m c}\,\left[\frac{{\cal E}}{2} - \frac{\alpha\, {\cal E}}{z_{+}t^{3/2}} \sin(k x-\omega t) \right. \nonumber \\ 
&& \left. \times \cos\left(\frac{\Omega\,t + \frac{\pi}{2}}{2}\right) \right];
\label{rho_EE} \\ 
\rho_{\mu}&\simeq& \frac{\hbar}{m c}\,\left[ \frac{{\cal E}}{2} -\frac{\alpha\, \mu}{z_{+}t^{3/2}} \cos(k x-\omega t) \right. \nonumber \\  
&&\left. \times \sin\left(\frac{\Omega\,t +\frac{\pi}{2}}{2}\right) \right], 
\label{rho_V}
\eeqa
where the frequency of the beats is given by the  ZBW frequency $\Omega=2\mu c=(2 m c^2/\hbar)$, and  $\omega={\cal E} c$ corresponds to the frequency of the point source. 
From Eqs.~(\ref{rho_EE}) and (\ref{rho_V}), 
we show that the ZBW effect emerges as a transient in the probability density, characterized by a superposition of two sinusoidal signals that exhibit a beating effect with a frequency $\Omega$. The amplitude of these transient oscillations decays as $t^{-3/2}$, and the stationary regime is reached as $t\rightarrow\infty$, and  $\rho\rightarrow {\cal E}$, as expected. 
The beating effect is accurately described by  Eqs.~(\ref{rho_EE}) and (\ref{rho_V}), as illustrated in Fig.~\ref{kg_zb_1}(a), where we show the  asymptotic density  $\rho_a$, and compare it with the exact density computed using  Eq.~(\ref{stepfinal}). 
Also, from Fig.~\ref{kg_zb_1}(b) and ~\ref{kg_zb_1}(c) we can appreciate that the amplitude of the oscillations for $\rho_{\cal E}$ are larger than those exhibited by $\rho_{\mu}$, which is consistent with the fact that $|{\cal E}|>\mu$, and verified by inspection of Eq.~(\ref{rho_V}). 
The exact value of the beating period for the cases discussed in Fig.~\ref{kg_zb_1}(a)
is $T=2\,\pi\,\Omega^{-1}=\pi$, as illustrated in the figure.
A similar behavior (not shown here) of the ZBW oscillations (beatings) was also observed in the Klein-tunneling regime, and the corresponding formulas for the density in the long-time regime can be derived in the same fashion as in the propagation case. 

Interestingly, the  observation and characterization of the  beating effect has been overlooked in studies that have addressed the problem of relativistic transients using different initial conditions \cite{mm52,moshrmf52,godoy16}. This is the case of the relativistic dynamics involving Klein-Gordon \cite{mm52} and Dirac  \cite{moshrmf52,godoy16} equations, that deal with  cut-off plane wave initial conditions within a quantum shutter setup. 
In fact, although not reported in Refs.~\onlinecite{mm52} and \onlinecite{moshrmf52}, we shall show in Sec.~\ref{relativistic_shutter_models} that the free Klein-Gordon  \cite{mm52} density and the Dirac \cite{moshrmf52} probability density within a quantum shutter setup exhibit similar beating effects associated to the ZBW phenomenon.

\subsection{Comparison with the free relativistic quantum shutter model} \label{relativistic_shutter_models}
\begin{figure}[H]
{\includegraphics[angle=0,width=3.3in]{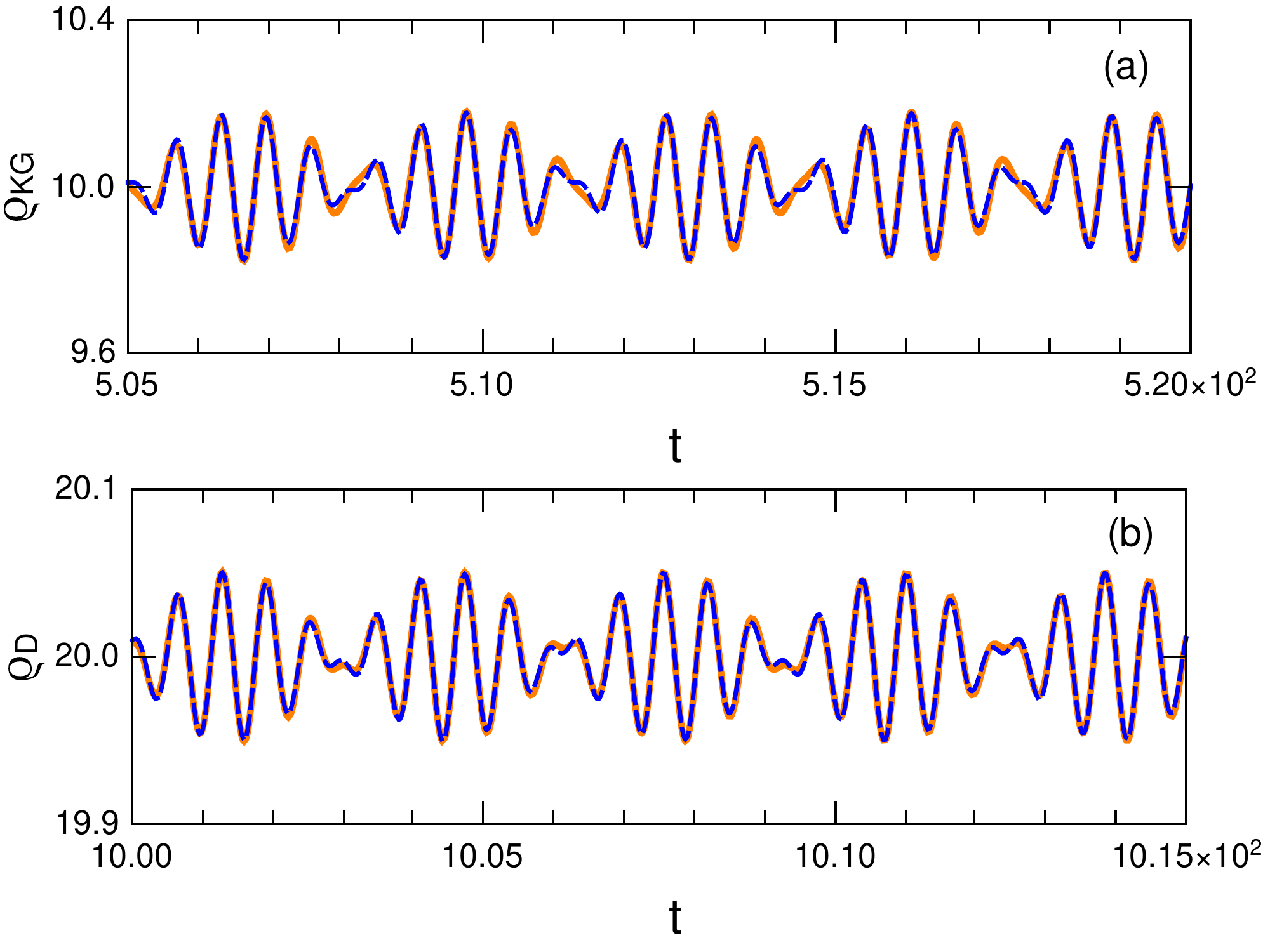}}
\caption{The ZBW effect in relativistic plane wave shutter models involving (a) free Klein-Gordon particles of spin $0$ and (b) Dirac particles of spin $1/2$. In both cases the ZBW effect appears as a series of {\it quantum beats} in the long-time behavior of particle density. In case (a) we plot the exact Klein-Gordon density using Eq.~(\ref{KG_libre_shutter}) (orange solid line), and the corresponding asymptotic density $\rho_{KG}$. In case (b) we plot the exact Dirac probability density $\rho=|\psi_1|^2+|\psi_2|^2$ (orange solid line) using Eq.~(\ref{Dirac_libre_shutter}), and the corresponding asymptotic probability density, $\rho_D$. In both cases we consider an energy $E=10.0$, and a fixed value of position $x=10.0$.}
\label{kg_d_zb_s_libre}
\end{figure}

We compare the results obtained for a \textit{point source model} with those of  systems involving a relativistic \textit{quantum shutter setup}. 
In general, while the quantum source problem involves an initial condition at $x=0$ of the form $\Psi(x=0,t)=e^{-i Ec t}\Theta(t)$, the relativistic quantum shutter model deals with cut-off plane waves at $t=0$ of the form  $\Psi(x,t=0)=e^{i k x}\Theta(-x)$ (Klein-Gordon case), and as a two-component spinor of the form $\boldsymbol{\Psi}(x,t=0)=[...,...]^T\,e^{i k x}\Theta(-x)$ (Dirac case). 
In particular, the Klein-Gordon  shutter problem \cite{moshrmf52} for a free particle of spin $0$,  deals with the solution of Eq.~(\ref{kg_equation}), with the initial condition at $t=0$ given by $\Psi_{KG}(x,0)=e^{ikx}\Theta(-x)$, and $[\partial_t \Psi_{KG}]_{t=0}=-i c E  e^{ikx}\Theta(-x)$. 
In our discussion of the Klein-Gordon and Dirac shutter models, we shall consider the free dispersion relation given by $E^2=k^2+\mu^2$, where $E$ is the energy in units of the reciprocal length.
The  time-dependent solution $\Psi_{KG}(x,t)$ is obtained by using the Fourier transform method \cite{moshrmf52},
\beqa
\psi_{KG}(x,t)&=&
\left[e^{i(k x-E ct)}+\frac{1}{2} J_0(\eta) 
-\sum\limits_{n=0}^\infty (\xi /iz)^nJ_n(\eta )\right] \nonumber\\
&\times& \Theta(t-x/c),
\label{KG_libre_shutter}
\eeqa
where $z=(E+k)/\mu$. We have found that the behavior of the above solution for $t \gg t_F$ is governed by 
$\psi_{KG}(x,t)\simeq e^{i(k x-E ct)}-\frac{1}{2} J_0(\eta)$, 
and following the procedure used in the point source case, we demonstrate that the free-density is given by  
$\rho_{KG}\simeq\rho_{KG,E}+ \rho_{KG,\mu}$ with
\beqa
\rho_{KG,E}&\simeq&\frac{\hbar}{m c}\,\left[\frac{E}{2} - \frac{\alpha_{KG}\, E}{t^{1/2}} \cos(k x-\omega t) \right. \nonumber \\ 
&& \left. \times \cos\left(\frac{\Omega\,t - \frac{\pi}{2}}{2}\right) \right];
\label{rho_EE_KG_f} \\ 
 \rho_{KG,\mu}&\simeq& \frac{\hbar}{m c}\,\left[ \frac{E}{2} +\frac{\alpha_{KG}\, \mu}{t^{1/2}} \sin(k x-\omega t) \right. \nonumber \\  
&&\left. \times \sin\left(\frac{\Omega\,t -\frac{\pi}{2}}{2}\right) \right], 
\label{rho_V_KG_F}
\eeqa
where  $\alpha_{KG}= (2\pi \mu c)^{-1/2}$.
The frequency of the \textit{quantum beats} is given by the  ZBW frequency $\Omega=2\mu c$, and  $\omega=E c$ corresponds to the frequency of the cut-off initial plane wave.
In Fig.~\ref{kg_d_zb_s_libre}(a) we plot the time-evolution of the exact  particle density in the long-time regime using Eq.~(\ref{KG_libre_shutter}), and we show the appearance of the \textit{quantum beat phenomena} associated to the ZBW effect. We also include for comparison the asymptotic probability  $\rho_{KG}$ and a perfect agreement is observed. 

The  quantum shutter problem for spin $1/2$ free particles in one-dimension \cite{moshrmf52,godoy16}, involves the solution of Dirac equation for a two-component spinor $\boldsymbol{\Psi}_{D}(x,t)=[\psi_1(x,t), \psi_2(x,t)]^T$, which satisfies,
\begin{equation}
\left[\frac{\sigma_0}{ic}\,\partial_t+\frac{\sigma_x}{i}\,\partial_x+\mu\, \sigma_z \right]\boldsymbol{\Psi}_{D}(x,t)=0,
\label{dirac_equation}
\end{equation}
where the helicity  is conserved, and $\sigma_i$ are the Pauli matrices. The corresponding cut-off initial condition at $t=0$ is,
\begin{equation}
\boldsymbol{\Psi}_{D}(x,0)=
\left[\begin{array}{c}
 1 \\
\frac{k}{E+\mu} 
\end{array}\right](E+\mu)^{1/2}\,e^{i k x}\,\Theta(-x).
\label{initial_cond_dirac}
\end{equation}
%
%
Moshinsky \cite{moshrmf52} showed that the solution of Eq.~(\ref{dirac_equation}) with the initial condition Eq.~(\ref{initial_cond_dirac}), is given by the spinor, %
\beqa
\boldsymbol{\Psi}_{D}(x,t)&=&\left\{ 
\left(\frac{\mu}{2z}\right)^{1/2}\Bigg( \left[\begin{array}{c}
 -1 \\
1 
\end{array}\right] J_0(\eta) +\left[\begin{array}{c}
 z+1 \\
z-1 
\end{array}\right]\Phi(x,t) \Bigg) \right\} \nonumber \\ &\times  &\Theta(t-x/c),
\label{Dirac_libre_shutter}
\eeqa
with $\Phi(x,t)$ defined as,
\beq
\Phi(x,t)=
e^{i(k x-E ct)} 
-\sum\limits_{n=1}^\infty (\xi /iz)^nJ_n(\eta).
\label{KG_libre_shutter_type}
\eeq
Note that in the Dirac solution Eq.~(\ref{Dirac_libre_shutter}), the spinor components are linear combination of Klein-Gordon type solutions similar to Eq.~(\ref{KG_libre_shutter}).
We have found that the long-time behavior of $\boldsymbol{\Psi}_{D}(x,t)$  can be accurately described by using $\Phi(x,t)\simeq e^{i(k x-E ct)} 
-(\xi /iz)J_1(\eta)$ in Eq.~(\ref{Dirac_libre_shutter}), which allows us to obtain a simple approximate formula for the Dirac  probability density 
$\rho_D=\boldsymbol{\Psi}_D^{\dag}\boldsymbol{\Psi}_D=\rho_{D,E}+\rho_{D,\mu}$, given by
\beqa
\rho_{D,E}&\simeq& E - \alpha_D \frac{\mu}{(E+k)}\frac{1}{t^{1/2}} 2 E \sin(k x-\omega t)  \nonumber \\ 
&& \times \cos\left(\frac{\Omega\,t+ \frac{\pi}{2}}{2}\right);
\label{rho_EE_D} \\ 
\rho_{D,\mu}&\simeq& E - \alpha_D \frac{\mu}{(E+k)}\frac{1}{t^{1/2}} \mu \cos(k x-\omega t)  \nonumber \\ 
&& \times \sin\left(\frac{\Omega\,t+ \frac{\pi}{2}}{2}\right), 
\label{rho_mu_D}
\eeqa
with  $\alpha_D= (8/\pi \mu c)^{1/2}$.
A plot of $\rho_D$ is shown in Fig.~\ref{kg_d_zb_s_libre}(b), where the \textit{beating phenomena} of the ZBW effect is observed. 
Notice also that for both Klein-Gordon and Dirac \textit{quantum shutter models}, 
the corresponding density and probability density exhibit similar features in the time-domain. 
Also, we found that the long-time behavior of the  densities for the \textit{shutter model} is governed by $t^{-1/2}$, while in the case of the \textit{point source model} the dynamics is governed by a $t^{-3/2}$ behavior.
Interestingly, these ideas have been recently applied to explore transients in two-dimensional Dirac systems like graphene \cite{cruz2019timediffraction}. The time-dependent solution obtained in Ref.~\onlinecite{cruz2019timediffraction} within a \textit{quantum shutter setup} is a two-component spinor whose components are linear combination of free Klein-Gordon solutions of the type given by Eqs.~ (\ref{simplifbis2}) or (\ref{KG_libre_shutter}).

%

%
We stress that the beating phenomenon observed in the long-time regime of the particle density, provides an alternative way to study the ZBW effect.
As discussed in Sec.~\ref{SZitter} the ZBW for Dirac-particles is not accessible by today experimental techniques since the effect corresponds to very high oscillation frequencies $\sim10^{21}$ Hz.  
Systems that mimic Dirac physics \cite{PhysRevLett.101.264303, PhysRevLett.105.143902,Gerri, LeBlanc_2013} have opened the possibility to measure the ZBW in more accessible frequency regimes.
For example, in graphene monolayers the dynamics of low-energy electron excitations are described by a Dirac-like equation.
We have recently shown that in these two-dimensional Dirac-like systems  \cite{cruz2019timediffraction}, the transient probability density exhibits quantum beats with a ZBW frequency in the terahertz regime.
The latter is accessible to experiments, and we suggest that the  quantum beat phenomena of ZBW can be probed  by means of current detection (instead of probability density) at ultrafast time-scales with high temporal resolution.

\section{Conclusions}\label{CONCLU}

We study the time-dependent features of quantum waves in the propagation and Klein-tunneling regimes of a potential step, by using an exact analytical solution to the Klein-Gordon equation with a point source initial condition. 
We show that the long-time behavior of the particle density exhibits a series of {\it quantum beats} characterized by the ZBW frequency, and that the amplitude of these transient oscillations decays as $t^{-3/2}$. 
We also show that the \textit{quantum beat} phenomenon is a robust effect that also manifest itself for free relativistic particles of spin $0$ and $1/$2, within a \textit{quantum shutter model} for Klein-Gordon \cite{mm52} and Dirac \cite{moshrmf52} equations.
We also find a time-domain where the density of the point source is characterized by a traveling main wavefront, which exhibits an oscillating pattern similar to the {\it diffraction in time} phenomenon \cite{mm52} observed in the non-relativistic case.
By measuring the relative positions of the main wavefronts associated to the step potential and free-case densities, we explored the features of time-delay, by implementing a criterion based on the energy difference, ${\cal E }$, between the point source and the potential step. 
We demonstrate that while in the propagation regime the density always exhibits a positive time-delay, in the Klein-tunneling regime the delay may be positive, negative or zero. The latter case corresponds to the so called super-Klein-tunneling configuration, where ${\cal E }$ equals to half the energy of the potential step. 

Finally, we argue that our alternative approach for investigating the ZBW effect opens the possibility for exploring this phenomenon in the probability density of Dirac fermions, by using different types of cut-off initial conditions. Although the ZBW in these relativistic systems is not accessible for experimental verification due to the high frequencies involved, we argue that 2D Dirac matter systems, like graphene in the low-energy regime, are ideal candidates for exploring the transient behavior of ZBW using  cut-off quantum waves \cite{cruz2019timediffraction}. This is  because in graphene \cite{RevModPhys.81.109} the Dirac fermions move with speeds hundreds of times smaller than the speed of light $c$, and hence this effect can manifest itself in the more accessible frequency range of terahertz.

\begin{acknowledgments}
The authors acknowledge support from UABC under Grant PFCE 2018. We also acknowledge useful discussions with R. Carrillo-Bastos. 
\end{acknowledgments}

\appendix

\section{Point source problem for the free-type case.}\label{PSB}

We include here for completeness the  procedure for obtaining the solution for the free-type case following the procedure of Ref. \onlinecite{dmrgv03}, which involves Laplace-transforming Eqs.~(\ref{kg_equation_2})  
and (\ref{CI_new}), using the standard definition 
\beq
\label{lapltrans}
\widetilde{\psi}(x;s)=\int_{0}^{\infty}\psi(x,t) \,e^{-st} dt.
\eeq
The Laplace transformed solution to  Eq.~(\ref{kg_equation_2}) reads, 
\beq
\widetilde{\psi}_0(x;s)=\alpha \,e^{-\sqrt{s^2+\mu^2 c^2}\,x/c},\qquad x\geq 0,  
\label{v4}
\eeq
and the Laplace transform of the initial condition [Eq. (\ref{CI_new})] is given by
\beq
\widetilde{\psi}_0(0;s)=\frac{1}{s+i c {\cal E}}.  
\label{v5}
\eeq
The matching  of Eqs. (\ref{v4}) and (\ref{v5}) at $x=0$, yields,
\beq
\widetilde{\psi}_0(x;s)=\frac{e^{-\sqrt{s^2+\mu^2 c^2}\,x/c}}{s+i c {\cal E}},\qquad x\geq 0.
\label{v6}
\eeq
The time-dependent solution $\psi_0(x,t)$ for $x\geq 0$ is obtained by performing the inverse Laplace transform of  $\widetilde{\psi}_0(x;s)$ 
by using the Bromwich inversion integral, which after some algebraic manipulations, yields,
\beq
\psi_0(x,t)=\psi_0^+(x,t)+\psi_0^-(x,t),
\label{bromwich_a}
\eeq
with 
\beq
\psi_0^{\pm}(x,t)=\frac{1}{2\pi i}\int\limits_{\gamma-i\infty }^{\gamma+i\infty }\frac 1 2\frac{(s-ic{\cal E})\,e^{-\sqrt{s^2+\mu^2 c^2}\,x/c}\,e^{st}}{\left(\sqrt{s^2+\mu^2 c^2}\pm i k c \right)\sqrt{s^2+\mu^2 c^2}}\,ds,   
\label{v7bisbis}
\eeq
where the integration is performed along a vertical line $Re[s]=\gamma$ in the complex $s$-plane, and all the singularities remain to the left-hand side of the line.
We evaluate (\ref{v7bisbis}) by using the change of variable %
$-iu=[(s^2+\mu^2c^2)^{1/2}+s]/\mu c$, which allows to write the integrals as,
\beq
\psi_0^{\pm}(x,t)=-\frac 1{2\pi i}\int\limits_{i\gamma'-\infty}^{i\gamma'+\infty}
{\cal F}^{\pm}(u)\,du,  
\label{v8}
\eeq
where ${\cal F}^{\pm}(u)$ is given by
\beqa
{\cal F}^{\pm}(u) =\frac{1}{2 u}\left[ \frac{u+z_{\pm}}{u-z_{\pm}}\right]\,e^{\frac{i}{2} \mu[u(x-c t)-u^{-1}(x+c t)]}, 
\label{v9bis}
\eeqa
with $z_{\pm}=({\cal E} \pm k)/\mu$. The  integration is  performed along the line ${\cal L}$ given by ${\rm Im}[u]=\gamma'$ in the complex $u$-plane, where all the singularities remain below ${\cal L}$.
We apply the Cauchy theorem to evaluate the integral for (\ref{v9bis}).  Let us first consider the case $x>ct$, where we close the integration path $\mathscr{L}$ from above with a large semicircle $\Gamma _1$ of radius $R$, forming the closed contour $\mathscr{C}_1$. The contribution along $\Gamma _1$ vanishes as $R\rightarrow \infty $, 
and since there are no poles enclosed by $\mathscr{C}_1$, 
the solution yields  $\psi_0(x,t)=0$ for $x>ct$. The second case corresponds to  $x<ct$, where we close the path $\mathscr{L}$ from below with a large semicircle $\Gamma _2$ of radius $R$, forming the closed contour $\mathscr{C}_2$, which also encloses two small circles $\mathscr{C}_0$, and $\mathscr{C}_{\pm}$ 
around an essential singularity at $u=0$, and simple poles at  $u=z_{\pm}$, respectively. 
The Cauchy integral formula yields,
\beqa
\psi_0^{\pm}(x,t)=\frac {1}{2\pi i}\left[\int_{\mathscr{C}_0} +\int_{\mathscr{C}_{\pm}}
\right]{\cal F}^{\pm}(u)\,du.
\label{solcuasifin}
\eeqa
The integrals corresponding to the contours $\mathscr{C}_{\pm}$ enclosing the simple poles are given by, 
\beq
\frac {1}{2\pi i}\int\limits_{\mathscr{C}_{\pm }}{\cal F}^{\pm}(u) \,du = 
e^{i[\pm k x-{\cal E} c t]},
\label{v11}
\end{equation}
and the contour integration about $\mathscr{C}_0$, which involves an essential singularity at $u=0$,  has already been obtained by  Moshinsky \cite{mm52} for the Klein-Gordon  quantum shutter problem. 
Thus, the resulting integral is, 
\beq
\frac 1{2\pi i}\int\limits_{\mathscr{C}_0} {\cal F}^{\pm}(u)\,du = -\sum\limits_{n=0}^\infty (\xi /iz_{\pm })^nJ_n(\eta )-\frac 1 2 J_0(\eta ),
\label{v15bis}
\eeq
with $\eta=\mu\,(c^2t^2-x^2)^{1/2}$, and  $\xi=[(ct+x)/(ct-x)]^{1/2}$. 
By substituting the results (\ref{v11}) and (\ref{v15bis}) into 
(\ref{solcuasifin}), the solution  $\psi_0(x,t)$ for $x>ct$ is obtained. Therefore, the  free-type relativistic solution is
\beq
\psi_0(x,t)=\left\{ 
\begin{array}{ll}
\psi_0^+(x,t)+\psi_0^-(x,t), & t>x/c; \\ 
0, & t<x/c,
\end{array}
\right.
\label{stepfinal2}
\eeq
where the solutions $\psi_0^{\pm}(x,t)$ are written as,
\beq
\psi_0^{\pm}(x,t)= e^{i[\pm kx-{\cal E}ct]}+\frac{1}{2}%
J_0(\eta ) -\sum\limits_{n=0}^\infty (\xi /iz_{\pm })^nJ_n(\eta ).
\label{simplif_bis}
\eeq
By substituting Eq.~(\ref{simplif_bis}) in Eq.~(\ref{psit_traslada}) we obtain the final solution for point source problem for a potential step, given by Eq.~(\ref{stepfinal}).

\section{Non-relativistic limit of $\psi(x,t)$} \label{LS}

The non-relativistic limit is obtained from 
the integral form of the  solutions $\psi^{\pm}_0(x,t)$ [Eq.~(\ref{v7bisbis})], that can be written with the help of the result $(s^2+\mu^2c^2)^{1/2}\simeq(\mu^2-i\beta/2)^{1/2}$ for a very large value of $c$, as 
\beqa
&&\psi^{(S)}_0(x,\pm k,t)\simeq \frac{1}{2\pi i}\\ 
&& \times\left[\int\limits_{\gamma-i\infty }^{\gamma+i\infty }\frac{1}{2} \frac{(s-ic{\cal E}) \,e^{-\sqrt{\mu^2-i\beta/2}}\,e^{st}}{c^2 \sqrt{\mu^2-i\beta/2} \left(\sqrt{\mu^2-i\beta/2}\mp ik \right)}\,ds \right], \nonumber \\
\label{sintrans}
\eeqa
where $\beta=(2m/\hbar)$. 
By performing the change o variable 
$-\sqrt{\mu^2+\beta/2i}=i\sqrt{\beta s^{\prime}}$ in Eq.~(\ref{sintrans}), we obtain,
\beqa
&&\psi^{(S)}_0(x,\pm k,t)\simeq e^{-i\mu ct} \frac{1}{2\pi i} \\ 
&&\times \left[\int\limits_{\gamma-i\infty }^{\gamma+i\infty } \frac{(2 s^{\prime}/c^2-i\beta/2-i{\cal E}/c) \,e^{i\sqrt{\beta is^{\prime}}}\,e^{2s^{\prime}t}}{-\sqrt{\beta is^{\prime}} \left(\sqrt{\beta is^{\prime}}\mp k \right)}\,\,ds^{\prime}  \right]. \nonumber
\eeqa
By taking the limit $c \rightarrow \infty$, $({\cal E}/c)\rightarrow 0$ and $(2s^{\prime}/c^2)\rightarrow 0$, the asymptotic solutions $\psi_S(x,\pm k,t)$ behave as, 
\beqa
&&\psi^{(S)}_0(x,\pm k,t) \simeq e^{-i\frac{\mu}{2} ct^{\prime}} \frac{1}{2\pi i} \nonumber \\ 
&&\times \left[\int\limits_{\gamma-i\infty }^{\gamma+i\infty } \frac{\left(\frac{i\beta}{2}\right) e^{i\sqrt{\beta is^{\prime}}}\,e^{s^{\prime}t^{\prime}}}{\sqrt{\beta is^{\prime}} \left(\sqrt{\beta is^{\prime}}\mp k \right)}\,ds^{\prime}  \right], 
\label{v7Schr}
\eeqa
where $t^{\prime}=2t$.
We can identify in Eq.~(\ref{v7Schr})  the integral representation of the Moshinsky's function 
\beqa
M(x,q,t)=\frac{1}{2\pi i} 
\int\limits_{\gamma-i\infty }^{\gamma+i\infty } \frac{\left(\frac{i\beta}{2}\right) e^{i\sqrt{\beta is^{\prime}}}\,e^{s^{\prime}t^{\prime}}}{\sqrt{\beta is^{\prime}} \left(\sqrt{\beta is^{\prime}}-q \right)}\,\,ds^{\prime}, \nonumber
\label{Moshint}
\eeqa
with $q=\pm k$. Therefore,
\beq
\psi^{(S)}_0(x,\pm k,t)\simeq e^{-i\frac{\mu}{2} ct}\, M(x,\pm k,t),
\eeq
and thus, $\psi_0(x,t)$ is given by
\beq
\psi^{(S)}_0(x,t)=e^{-i\frac{\mu}{2} ct}\, [M(x,k,t)+M(x,-k,t)].
\label{Schro0Sol}
\eeq
From Eq.~(\ref{psit_traslada}), we finally obtain the no-relativistic limit of the Klein-Gordon solution for a potential step
\beq
\psi^{(S)}(x,t)=e^{-i\frac{\mu}{2} ct} \left\{\left[M(x,k,t)+M(x,-k,t)\right]\,e^{-iU_r t/\hbar}\right\},
\label{SchroSol}
\eeq
which coincides with the analytical solution of  Schr\"odinger's equation for a point source initial condition for a step potential barrier \cite{jvrrsss02}.


\end{document}